\documentclass[12pt]{article}
\usepackage{graphics}
\begin{document}
\bigskip \bigskip \bigskip
\centerline{\large \bf Exponent 1/4 from the mean field approximation}

\bigskip \bigskip
\centerline{\bf Aleksey Nudelman}
\medskip
\centerline{Department of Physics}
\centerline{University of California}
\centerline{Santa Barbara, CA  93106-9530 U.S.A.}
\medskip
\centerline{email: anudel@physics.ucsb.edu}
\bigskip \bigskip                           
\begin{abstract}
	We consider the magnetization of the three-dimensional
ferromagnetic near the critical temperature in the mean field
approximation. It is found that for some directions of easy magnetization
one obtains exponent 1/4 instead of the usual 1/2.

KEY WORDS: Critical exponent. Spontaneous magnetization. Mean field
approximation. Directions of easy magnetization.
\end{abstract}

	Our model in three dimensions is as follows: we consider
the cubic
crystal lattice with spins situated at vertexes and introduce a coordinate
system along the edges of the cell. Then directions of the spin can be
specified relative to this basis as a unit
vector $(a_1,a_2,a_3)$. To make
our results coordinate-independent we induce the symmetry transformation
around
a vertex. We apply three 90--degree rotations {\bf$C^4$} around
coordinate
axes 1, 2, 3 and make one of the planes 12, 13, 23 a mirror plane. Doing
this
we generate from any given spin orientation 48 other directions
$(a_1,a_2,a_3)$ obtained by 3! permutations of $a_1,a_2,a_3$ times 8
permutations of the sign. (Compare it with the conventional mean field
approach where
spins are either up or down.)
		Magnetization per unit cell $M$ in the direction
of the external magnetic field $H$ is defined as 
\begin{equation}
M=\frac{\sum_{i=1}^{48}{p_i}exp[\beta H {p_i}]}{Z[H]}
\end{equation}
where $p_i$ is a projection of the spin on the direction of the
magnetization, $\beta$ is the inverse temperature and $i$ specifies a
given
spin
orientation generated by the symmetry group. The statistical sum is
defined as
\begin{equation}
Z[H]=\sum_{i=1}^{48}exp[\beta H{p_i}]
\end{equation}
In the mean field approximation the magnetization "eats its own tail"
\cite{wanier} i.e.
\begin{equation}
M=\frac{\sum_{i=1}^{48}{p_i}exp[\beta M {p_i}]}{Z[M]}\label{mag} 
\end{equation}
This nonlinear equation can be solved near the critical temperature where
the spontaneous magnetization becomes small. Given the fact that the right
hand side of eq.\ref{mag} is an odd function of $M$, the first three terms
in
the Taylor expansion are proportional to  $M$, $M^3$, $M^5$
respectively. If we are to limit ourselves to the first term
of the Taylor expansion we would get an expression for the critical
temperature in terms of $p$ and $k_B$ after $M$ being factored out from
both sides of eq.\ref{mag}. Then eq.\ref{mag} would look like 
\begin{equation}
1\approx\frac{T}{T_c} + f(a_1,a_2,a_3) \frac{M^2}{T^2} +
g(a_1,a_2,a_3)\frac{M^4}{T^4}
\label{ta}
\end{equation}
where
\newline
\(\begin{array}{r}
f(a_1,a_2,a_3)={p^2}/{6{k_B}^2}[(4{a_2}^2{a_3}^2-{a_1}^4)({a_2}^4+{a_3}^4)\\
+(4{a_1}^2{a_3}^2-{a_2}^4)({a_1}^4+{a_3}^4)\\
+(4{a_1}^2{a_2}^2-{a_3}^4)({a_1}^4+{a_2}^4)]
\end{array}\)
\newline
and $g(a_1,a_2,a_3)$ is to be specified later. Please note that
we
expressed the contributions of all 48 directions of the spins in terms of
the
first direction, $(a_1,a_2,a_3)$, from which they were generated by the
symmetry
group. Our focus now is on the different classes of the spins generated by
$(a_1,a_2,a_3)$. For all practical purposes  we can neglect the third term
in eq.\ref{ta} relative to the second term and the magnetization becomes
\begin{equation}
M\approx\frac{{k_B}T}{p} \sqrt{\frac{6(1-T/{T_c})}{f(a_1,a_2,a_3)}}
\end{equation} 
where $T_c=\frac{p^2}{3k_B}$.
By the substitution 
\begin{equation}
x=1-{a_3}^2={a_1}^2+{a_2}^2
\label{st}
\end{equation}
\begin{equation}
\nonumber
a=\frac{{a_1}^2{a_2}^2}{({a_1}^2+{a_2}^2)^2}
\label{su}
\end{equation}
we obtain
\begin{equation}
M\approx\frac{k_BT}{p}\sqrt{\frac{3(1-T/T_c)}{x[-5x^3(a-1)^2+
10x^2(1-a)-x(7-2a)+2]}}
\label{ma}
\end{equation}
This procedure would fail if $f(a_1,a_2,a_3)$ vanishes. If
this happens we may no longer neglect the third term in eq.\ref{ta} 
and one obtains $M \sim (T_c-T)^{1/4}$ instead of eq. \ref{ma}.
Let us obtain the range of $a$ corresponding to the exponent 
1/4.
	It follows from eqs. \ref{st}-\ref{su} that for $x \in [0,1]$ and
$a \in
[0,1/4]$
$f(a_1,a_2,a_3)$ vanishes iff $x=0$ or $x=1$. 
 Please note that $x=0$ corresponds to the orientation of
the spins in the faces of the unit cell. Thus $x=0$ is a
particular case of $x=1$. To illustrate this simple case we have plotted
the dependence of the magnetization on the temperature in figure 1.
                         
\begin{figure}[h]
\begin{center}
	\scalebox{0.3}{\includegraphics{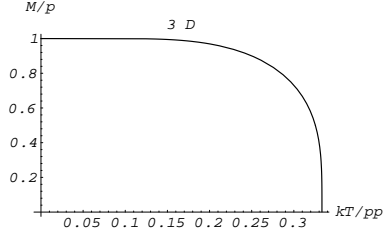}}
\caption{The dependence of the magnetization on the temperature when spins 
are aligned along the edges of the three dimensional cubic cell.}
\end{center}
\end{figure}
	
	It is interesting to see if the exponent 1/4 obtained above is
consistent with the mean field approximation. Let us introduce the two-particle
correlation
function $<{\overrightarrow{p}_1}{ \overrightarrow{p}_2}>/<{
\overrightarrow{p}^2}>$ \newline
($<p>$ means statistical average of $p$
and D
is the dimension of the cubic lattice) for the simple case $x=0$
\begin{equation}
<{\overrightarrow{p}_1}{\overrightarrow{p}_2}>=\frac{cosh[2{Mp}/{{k_B}T}]-1}{1+cosh[2y]+
4(D-1)cosh[{Mp}/{{k_B}T}]+2(D-1)^2}
\end{equation}
\begin{equation}
<\overrightarrow{p}^2>=1- \frac{D-1}{cosh[{Mp}/{{k_B}T}] + D-1}
\end{equation}
\begin{figure}[h]
\begin{center}
	\scalebox{0.4}{\includegraphics{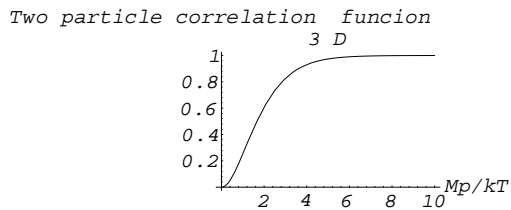}}
\caption{The dependence of the two-particle correlation function
on $Mp/{{k_B}T}$ 
for the 3D cubic crystal with the critical exponent 1/4}
\end{center}
\end{figure}
As it is seen from figure 2, the exponent 1/4 does not lead to any
qualitative change in the behavior of the correlation function as compared
to the well studied 1/2 case.
Let us note that the mean field approximation is derived under
the assumption
that the correlation
between spins at the different lattice sites is small. As it is seen from
the
graph the correlation function remains small as long as the magnetization
is close to the critical temperature. This behavior is to be contrasted
with
that of the 4D cubic crystal which we consider below.

To simplify calculations we consider the four-dimensional
cubic lattice.
A spin may align itself in 8 directions along the cell's
edges. The magnetization
\begin{equation}
M=p\frac{sinh[Mp/{{k_B}T}]}{cosh[Mp/{{k_B}T}]+3}
\label{sq}
\end{equation}
 is no longer small near the critical temperature as can be seen
from
figure 3.
\begin{figure}
\begin{center}
	\scalebox{0.3}{\includegraphics{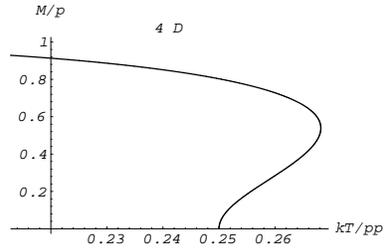}}   
\caption{The dependence of the magnetization eq.\ref{sq} on the
dimensionless
temperature.}
\end{center}
\end{figure}
The $4D$ case (as well as any higher dimensional lattice) is not described
well by the mean field approximation because, as seen from figure 4, the
two-particle correlation function is not small near the critical
temperature.
\begin{figure}
\begin{center}
	\scalebox{0.4}{\includegraphics{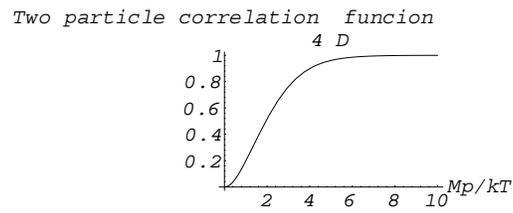}}
\caption{The dependence of the two-particle correlation function on
$Mp/{{k_B}T}$ for the 4D cubic crystal.}
\end{center}
\end{figure}

In conclusion we show that the exponent 1/4 is not possible in the
two-dimensional case. (It is also not possible to get
1/4 for
the one-dimensional crystal as well but the calculations are so trivial
that
we do not show them here.) In the two-dimensional crystal we define
the direction
of easy magnetization as $(a_1,a_2)$. There are 8 directions of the spin
relative to $M$ generated by the inversion around  the coordinate origin
and by the 90-degree rotations in the plane of the crystal. Near the
critical
temperature, after some algebra, one obtains the following expression for
the magnetization:
\begin{equation}
M \approx \frac{2Tk_B}{p} \sqrt{\frac{1-T/T_c}{1-\frac{1}{3}(1+
({a_1}^2-{a_2}^2)^4 + (2{a_1}{a_2})^4)}}
\end{equation}	
where $T_c = \frac{p^2}{2k_B}$ and $p$ is a spin magnetic moment. If we
substitute $a_1 =cos\phi$ 
	$a_2=sin\phi$
then $M \approx \frac{2Tk_B}{p}
\sqrt \frac{1-T/T_c}{1+\frac{sin^2 4\phi}{2}}$
\begin{equation}
1+\frac{sin^2 4\phi}{2}\neq0
\end{equation}
and therefore
\begin{equation}
M \sim
\sqrt{(T-T_c)} 
\end{equation}
for all directions of easy magnetization.
\\
\newline

{\bf Acknowledgements}

It is a pleasure to thank S. I. Ben-Abraham, Walter Poetz, Andreas X
for discussions and Ms. Y. C. who worked hard on my grammar but refused to
be
acknowledged.

\end{document}